# Isofrequency spin-wave imaging using color center magnetometry for magnon spintronics

Samuel Mañas-Valero,[1*] Yasmin C. Doedes,[1] Artem Bondarenko,[1] Michael Borst,[1] Samer Kurdi,[2] Thomas Poirier,[3] James H. Edgar,[3] Vincent Jacques,[4] Yaroslav M. Blanter,[1] Toeno van der Sar[1*]

1 Department of Quantum Nanoscience, Kavli Institute of Nanoscience, Delft University of Technology, Delft 2628CJ, the Netherlands

2 Institute of Photonics and Quantum Sciences, SUPA, Heriot-Watt University, Edinburgh EH14 4AS, United Kingdom

3 Tim Taylor Department of Chemical Engineering, Kansas State University, Kansas 66506, USA

4 Laboratoire Charles Coulomb, Université de Montpellier and CNRS, 34095 Montpellier, France

**Corresponding Authors**

Dr. Samuel Mañas-Valero (S.ManasValero@tudelft.nl)
Prof. Toeno van der Sar (T.vanderSar@tudelft.nl)

**Abstract**

**Magnon spintronics aims to harness spin waves in magnetic films for information technologies. Color center magnetometry is a promising tool for imaging spin waves, using electronic spins associated with atomic defects in solid-state materials as sensors. However, two main limitations persist: the magnetic fields required for spin-wave control detune the sensor-spin detection frequency, and this frequency is further restricted by the color center nature. Here, we overcome these limitations by decoupling the sensor spins from the spin-wave control fields –selecting color centers with intrinsic anisotropy axes orthogonal to the film magnetization– and by using color centers in diamond and hexagonal boron nitride to operate at complementary frequencies. We demonstrate isofrequency imaging of field-controlled spin waves in a magnetic half-plane and show how intrinsic magnetic anisotropies trigger bistable spin textures that govern spin-wave transport at device edges. Our results establish color center magnetometry as a versatile tool for advancing spin-wave technologies.**

## Introduction

Quantum sensing based on color centers in solid-state hosts, such as the nitrogen-vacancy (NV) center in diamond and the boron vacancy ($V_B$) center in hexagonal boron nitride (hBN), employs the electronic spin associated with the center for interacting with the environment and unveiling its physical properties (**Fig. 1a-b**).[1] The operation of these solid-state spin sensors relies on their optically detectable electron spin resonance (ESR) spectrum, which is sensitive to a range of physical quantities such as magnetic fields, electric fields, temperature, and pressure.[2–6] Combined with the ability to place the spin sensors within nanometers from a material of interest, as required for high-resolution spatial imaging, this sensitivity has enabled applications in fields ranging from biology and (bio-)chemistry to condensed matter physics and geoscience.[7–9]

Because of its high sensitivity, spatial resolution, and ability to image both static and dynamic magnetic fields, spin-based magnetometry is particularly well suited for high-resolution local probing of magnetic materials, exemplified by recent experiments probing atomically-thin magnets[10,11] and imaging spin-wave dynamics in magnetic thin films.[12–15] Such spin waves - collective wave-like spin excitations with quasi-particle excitations called magnons - play a central role in the thermodynamics of magnetic materials and are promising as signal carriers for information devices due to their low intrinsic damping, non-reciprocal transport properties, micrometer wavelengths at microwave frequencies, and strong interactions that enable signal transduction.[16–18] A central prospect of magnon spintronics is to harness spin waves for microwave control in chip-scale devices. Developing such devices requires high-resolution, high-sensitivity sensors that can image both the spin-wave dynamics and the underlying spin textures in thin-film magnets.[19]

Compared to other spin-wave imaging techniques, spin-sensor-based spin-wave imaging stands out because it detects spin waves by their magnetic stray fields.[12,13] This detection mechanism provides the ability to image spin waves underneath optically opaque materials and to image the static spin textures or electrical currents with which the waves can interact.[15,20] However, a key challenge for spin-wave imaging using spin-based magnetometry is the resonance requirement between the sensor-spin frequency and the frequency of the spin wave:[12,13,15,20] although the sensor spin can be tuned to a target frequency by a magnetic bias field, such bias field generally also couples to and thereby changes the spin-wave spectrum of the target sample. As such, the coupling hampers the ability to image spin waves at target frequencies, such as the operating frequency of spin-wave devices.

Here we overcome this challenge by decoupling to first order the control of the sensor-spin frequency from that of the spin-wave dynamics in a magnetic thin film. The key concept is to orient the intrinsic anisotropy axis of the sensor spins along the hard axis of the magnetic sample. This geometry decouples the magnetic-field control of the sensor-spin frequency from the magnetic system, enabling independent tuning of the two systems.

We use this new technique to demonstrate isofrequency imaging of field-controlled spin waves in a thin-film permalloy magnet, using spins in both diamond and hexagonal boron-nitride to access complementary frequency ranges. By tuning both the angle and the magnitude of the magnetic bias field, we control the spin-wave propagation in the inhomogeneous spin textures at the edge of a permalloy magnetic half-plane, representing a case study of spin-wave transport in spin textures that arise at the edges of lithographically patterned magnetic devices. We tune the spin-wave length by a factor six by applying a magnetic bias field that compensates a small in-plane anisotropy, and then harness this anisotropy to deterministically nucleate bistable, curling spin textures at the magnetic edge that enable different spin-wave transport regimes. We extract the magnetic curling length (the length over which the spins reorient) and reproduce the observed bistable textures via simulations. Our static- and dynamic-field images reveal the important role of the spin textures underlying the spin-wave transport in patterned magnetic films — central elements for spin wave devices — and pinpoint spin-based magnetometry based on complementary color centers as a versatile technique for its study.

## Results

### First-order decoupling of the spin sensors and spin waves

To image spin waves, we use sensor spins associated with NV centers in diamond and $V_B$ centers in boron nitride (**Fig. 1a,b**). Crucially, both the NV and the $V_B$ color center have an effective spin of magnitude $S = 1$.[7,21] As such, the crystallographic symmetry of the defect endows the spin with an intrinsic anisotropy axis. For the NV center, this axis is oriented along the line connecting the nitrogen and vacancy sites (**Fig. 1a**).[7] For the $V_B$ center, this axis is perpendicular to the layers of the hBN crystal (**Fig. 1b**).[22] The anisotropy causes the spin energy levels to be split even at zero magnetic field (by $D$ = 2.87 GHz for the NV and $D$ = 3.5 GHz for the $V_B$) and renders the ESR frequencies first-order (second-order) sensitive to magnetic fields that are oriented parallel (perpendicular) to the anisotropy axis (**Fig. 1c,d** and **Methods**). Similarly, the shape anisotropy of a thin-film magnet renders its spin-wave spectrum first-order sensitive (insensitive) to in-plane (out-of-plane) magnetic fields (**Fig. 1c,d** and **Methods**). As such, by selecting sensor spins with anisotropy axis perpendicular to the magnetic plane, we gain independent magnetic-field control of the two systems, enabling isofrequency imaging of field-controlled spin waves.

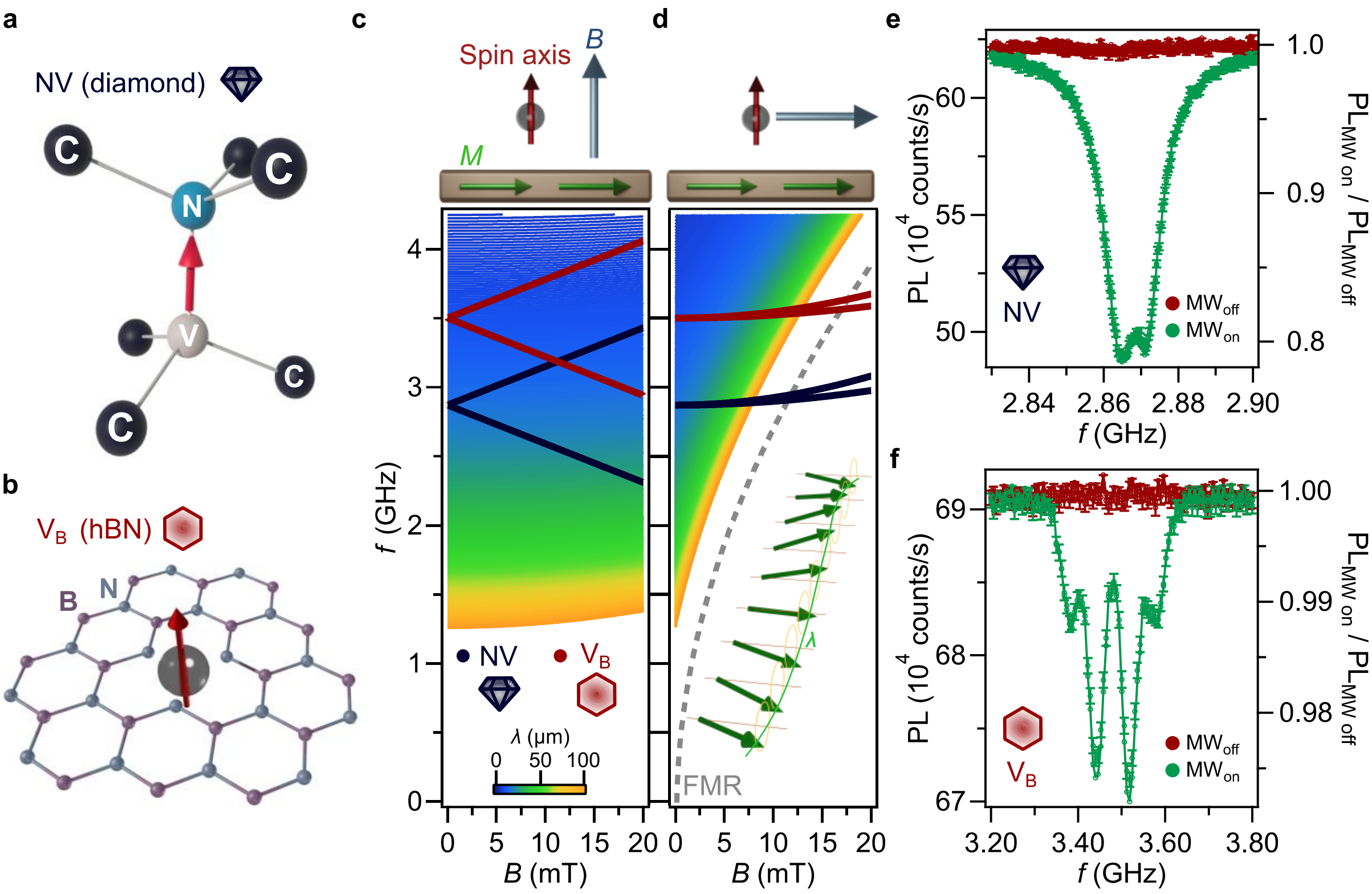

**Figure 1.- Spin wave imaging using solid-state spin sensors that are first-order decoupled from the spin-wave spectrum. a, b)** Atomic structure of the nitrogen-vacancy (NV) center in the diamond carbon (C) lattice (a) and of the boron vacancy ($V_B$) in a hexagonal boron nitride layer (b). **c, d)** Dispersion of the spin sensors and the spin waves. Blue and red lines: Magnetic field dependence of the electron spin resonance (ESR) frequencies of the NV (zero-field splitting $D$ = 2.87 GHz) and $V_B$ (D = 3.5 GHz) center, respectively. Color map: spin-wave length λ. The red, blue and green arrows above the graphs indicate the spin anisotropy axis, externally applied magnetic field, and easy-plane magnetization respectively. Grey dashed line: the ferromagnetic resonance (FMR) according to Kittel's law. In (c), the magnetic field $B$ is applied *along* the spin anisotropy axis, linearly splitting the ESR frequencies $f_-$ and $f_+$ while leaving the spin-wave length λ unaffected to first order. In (d), $B$ is applied *perpendicular* to the spin anisotropy axis, rendering the ESR frequencies decoupled from $B$ to first order while strongly coupling to the spin-wave length. See **Methods** for the mathematical derivation. Inset in (d): sketch illustrating the spin-wave length λ. **e-f)** Optically detected ESR spectrum near zero applied magnetic field for the NV (e) and $V_B$ (f) spins. Monitoring the ESR contrast, defined as C = 1 – $PL_{MWon}/PL_{MWoff}$, where $PL_{MWon}$ ($PL_{MWoff}$) is the photoluminescence (PL) under (in the absence of) microwave driving, enables spin wave detection. The four-peak structure of the $V_B$ spectrum is associated to hyperfine coupling to nearby $^{15}N$ spins.[21]

### Spin wave imaging using color centers

Our spin wave imaging method is based on the spin-dependent photoluminescence of the color centers, which decreases under microwave driving at the ESR frequency (**Fig. 1e,f**).[7,23] By placing the sensor spins close to the sample surface **(Fig. 2a,b**), the magnetic stray fields generated by nearby spin waves directly drive the ESR transitions of the sensors,[14,24–26] enabling spin wave imaging by spatially mapping the ESR contrast **(Fig. 2c)**.[12,13,15,27] To achieve proximity to the sample, we place a flake of hBN on top of the magnetic film and use a single NV spin in a scanning diamond tip (**Fig. 2a).** The range of detectable frequencies is set by the width of the ESR spectra of the spin sensors. As such, the ESR spectrum of the $V_B$ center provides a broader range of detectable frequencies than the NV center, [28] at the cost of a reduced sensitivity.

**Isofrequency imaging of field-controlled spin waves**

We demonstrate that by using an orthogonal configuration between the sensor spins and the spin-wave spectrum enables isofrequency imaging of field-controlled spin waves (**Fig. 2)**. Representative images of NV and $V_B$ resonant spin waves under a magnetic bias field of $B$ = 1.6 mT applied along the microstrip (so-called Damon-Eshbach configuration) are shown in **Fig. 2c-e**. By changing the in-plane magnetic field strength, we control the spin-wave length while the first-order decoupling enables keeping the imaging frequencies fixed to the zero-field splitting of the NV and $V_B$ center (**Fig. 2f-h**, left panels). Our experimental observations are captured well by computing the magnetic stray field of a spin wave traveling perpendicularly to the static magnetization (**Fig. 2f-h**, right panels).[12] Although the measurable field range is limited by the photoluminescence quenching of the color centers under large off-axis fields,[29] we do not reach such fields in these measurements. Instead, the range is limited by the spin-wave band being pushed above the ESR detection frequencies by the applied field (**Fig. 1d**). This observation emphasizes the relevance of studying off-axis fields in color centers.[29,30] We highlight that the spin waves are also visible underneath the metal as a result of the magnetic nature of the detection method.[20] These results demonstrate the ability to decouple the magnetic-field control of the spin-wave spectrum from the detection frequency of the color centers, which can be expanded to other frequencies by employing additional color centers such as carbon-related defects in boron nitride of silicon vacancies in silicon carbide.[30,31]

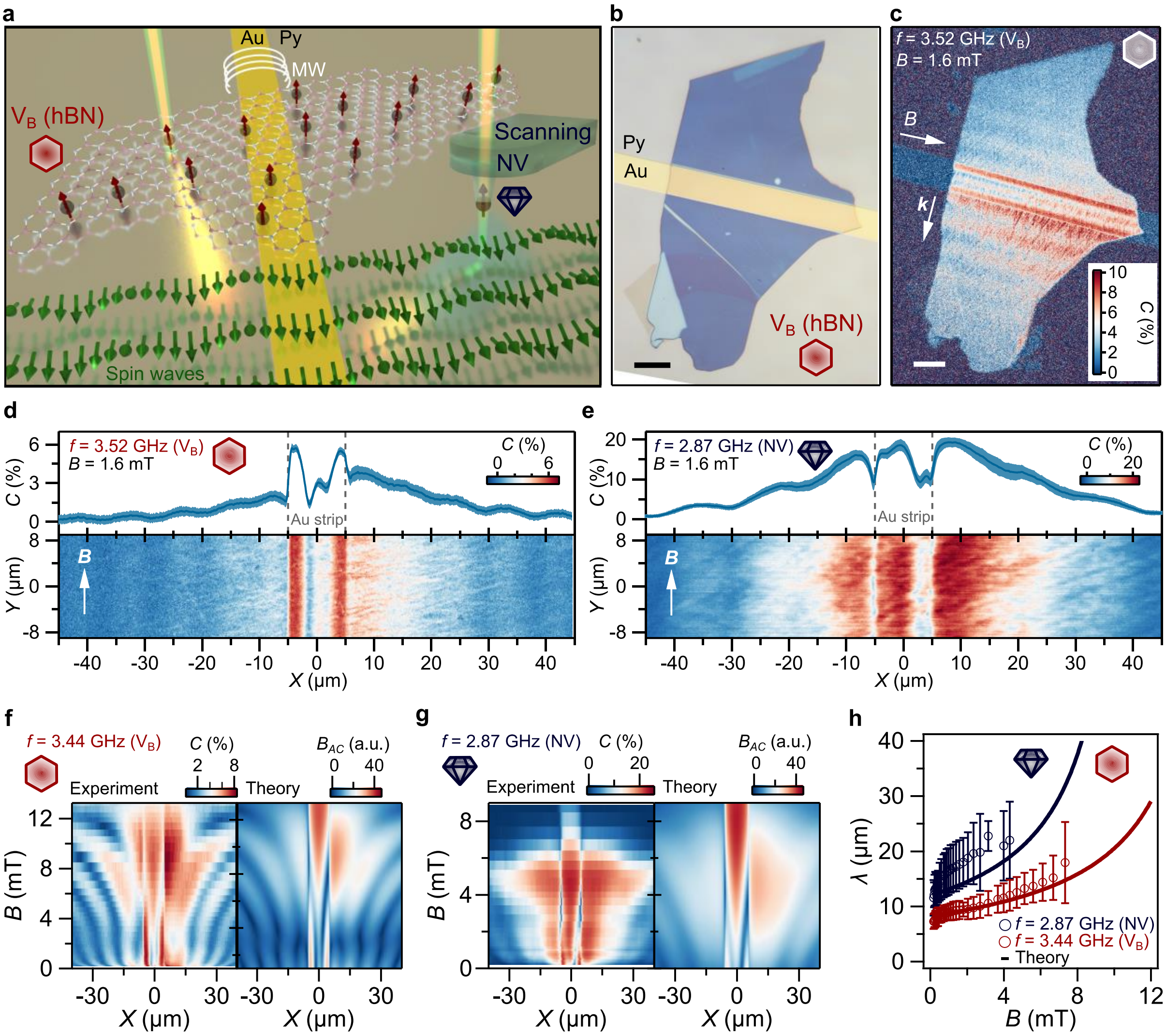

**Figure 2.- Isofrequency imaging of field-controlled spin waves using color centers. a)** Sketch of the experiment: a microwave (MW) current in a 80-nm-thick gold (Au) microstrip excites spin waves (green arrows) in a 55-nm-thick permalloy (Py) film (**Methods**). We detect the waves using both an ensemble of $V_B$ sensor spins in a hexagonal boron nitride flake placed on top of the sample and a single NV sensor spin in a diamond tip scanned across the sample. **b**) Optical image showing the $V_B$ -containing boron-nitride flake on top of the Py sample and Au microstrip. Scale bar: 10 µm. **c**) ESR contrast map showing spin waves travelling perpendicular (Damon-Eshbach spin waves) to the microstrip with wavevector **k**. A wavelength of 7.9(8) µm is extracted by fast Fourier transformation (FFT, **Supplementary Section 1.4**). An in-plane field $B$ = 1.6 mT along the microstrip and a $V_B$ -resonant drive frequency of 3.52 GHz yields Damon-Eshbach spin waves. Scale bar: 10 µm. **d-e**) Spin-wave maps (bottom panel) and their average along Y (top panel) for $B$ = 1.6 mT applied in-plane along the microstrip (white arrows). The drive frequency is 3.55 GHz for the $V_B$ measurement in (d) and 2.87 GHz for the NV measurement in (e). The FFT-extracted wavelengths are 8.2(8) and 10.6(9) µm, respectively. The gold microstrip is located between X ∈ [-5, 5] µm. Shaded area in top panels: ±1 standard deviation. **f-g**) Isofrequency imaging of field-controlled spin waves. Measured (left panels) and calculated (right panels) 1D spin-wave maps vs. $B$. The field $B$ is applied in-plane along the microstrip. Drive frequency: 3.44 GHz for the $V_B$ measurement in (f) and 2.87 GHz for the NV measurement in (g). An external MW reference field is added in f (see **Methods**). **h)** Spin-wave length vs. field strength $B$. The 1D profiles shown in panels (f) and (g) are obtained from averaging the 2D spatial contrast maps presented in **Supplementary Section 2.1**. The spin-wave length shown in (h) is extracted using a fast Fourier transform (FFT) applied to these 2D maps, as described in **Supplementary Section 1.4**.

**Spin wave imaging in a magnetic half-plane with competing magnetic interactions**

Realizing spin-wave circuits and devices relies on patterned magnetic films. Such patterning generally leads to inhomogeneous spin textures such as domain walls or curling textures at the device edges.[32–34] Equipped with a strategy for isofrequency imaging of field-controlled spin waves, we now consider a magnetic half-plane as a simple device model (**Fig. 3a-b**). We show how the inhomogeneous spin textures affect the spin-wave transport and that field-balancing the magnetic anisotropy enables controlling the local spin-wave length .[35–38]

Using a single NV spin in a scanning diamond tip (**Fig. 3a**), we image spin waves at the edge of the plane as a function of the angle φ of an in-plane magnetic field (**Fig. 3c**). We employ an NV spin due to its higher sensitivity to the static magnetic fields generated by the sample. When the field is oriented along the edge (φ = 90°), we observe spin waves with a spatially uniform wavelength. This is expected from the shape anisotropy introduced by the edge, which favors the underlying magnetization to be uniform and oriented along the edge. Reversing the field to φ = 270°, we observe the same wavelength but different amplitudes, attributed to the non-reciprocity of this Damon-Eshbach configuration.[39,40] Rotating the field from φ = 90° towards φ = 0° (**Fig. 3c-e**), we observe that the spin-wave length initially stays constant and then decreases rapidly by a factor ~6 towards the few-micrometer regime. Remarkably, at φ = 0°, the spin-wave length varies spatially (**Fig. 3c** and **Supplementary Section 2.2.8**), decreasing with increasing distance to the edge before becoming constant at $X \approx 20$ μm. We attribute this change to an underlying, inhomogeneous spin texture near the film edge, which we corroborate below using static-field imaging and micromagnetic simulations.

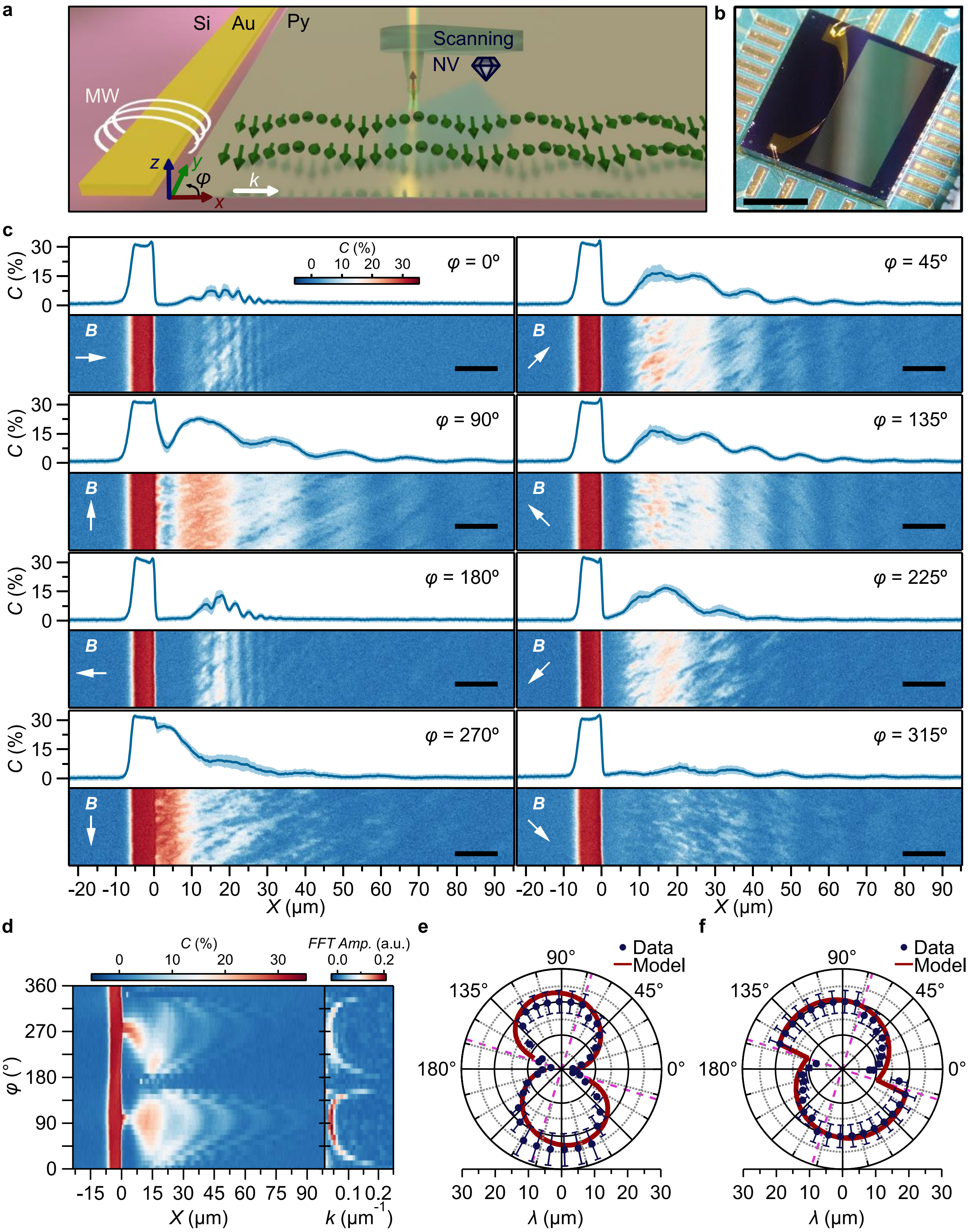

**Figure 3. Isofrequency imaging of anisotropy-governed spin waves under angular field control. a)** Sketch of the experiment. A microwave (MW) current in a gold (Au, yellow) microstrip excites spin waves (green arrows) with wavevector k (white arrow) in a 50-nm-thick permalloy (Py) half-plane. A single NV spin (lattice structure shown in inset) in a scanning diamond tip images spin waves resonant with the NV ESR frequency by detecting their microwave magnetic stray fields. **b)** Optical image of the magnetic half-plane. Scale bar: 2 mm. **c)** Spatial spin-wave maps (bottom panels) and their average along the vertical direction (top panels) for eight different angles φ (in-plane angle from *x* towards *y* as defined in a) of the applied magnetic field ($B$ = 0.98 mT). Scale bar: 10 µm. **d)** Spin-wave maps *vs.* field angle φ

(left panel) and the corresponding spin-wave number $k$ extracted via fast Fourier transformation (FFT, right panel) at $B$ = 0.98 mT. **e-f)** Spin wave length *vs.* in-plane field angle φ at $B$ = 0.98 mT (**e**) and $B$ = 0.39 mT (**f**). The sudden change in (f) signifies the angle at which the applied field is perpendicular to the anisotropy axis, enabling us to extract an anisotropy angle of $\varphi_K$ = 73(2)° (highlighted with pink dashed line). Red lines: calculation based on the Landau-Lifshitz equation for an anisotropy field of $B_K$ =1 mT that is oriented at $\varphi_K$ = 73°. The corresponding spatial spin-wave maps are shown in **Supplementary Section 2.2**.

To further investigate the role of anisotropy, we extract the average spin-wave length via Fourier transformation from the spatial spin-wave maps (**Fig. 3d**). Doing so at both $B$ = 0.98 mT (**Fig. 3e**) and $B$ = 0.39 mT (**Fig. 3f**), we observe a strikingly asymmetric behaviour and a jump in the spin-wave length for φ = 165° and 345° (most clearly visible in **Fig. 3f**), even though our sample is mirror-symmetric w.r.t. the x-axis. Using the LLG equation to calculate the spin-wave length in a uniform permalloy film with a 1 mT uniaxial in-plane anisotropy oriented along φ = 75° (**Methods** and **Supplementary Section 2.2.9)**, we are able to accurately reproduce both the asymmetry and the jump (red lines in **Fig. 3e-f**), with the jump caused by the magnetization switch that occurs when the field is perpendicular to the anisotropy axis. We conclude that in addition to the shape anisotropy introduced by the edge, our film has a small in-plane anisotropy that breaks the symmetry of the system, as corroborated by the different spin wave amplitudes between φ = 0° and φ = 180° in **Fig. 3c** and **Supplementary Section 2.2.8**. These differences are attributed to an underlying variation in the curling magnetic configuration present in each case. Such small anisotropies are typical in permalloy films grown by various methods.[41,42]

We demonstrate that balancing the magnetic anisotropy by a submillitesla magnetic field applied in target directions enables a high degree of control of the spin-wave length. When the field is applied along the edge (φ = 90°, **Fig. 4a**), there is only a weak dispersion, in contrast with the highly dispersive response for φ = 180° (**Fig. 4b**). In this case, the system is in a curling magnetic state, where the spin-wave length varies spatially, as shown in **Supplementary Section 2.2.8**. This configuration does not correspond to a standard backward-volume geometry due to the non-uniform, spatially rotating magnetization. An intuitive understanding of the field dependence is as follows: at lower applied fields, the degree of curling is reduced, resulting in a magnetization orientation that is closer to being perpendicular to the *k*-vector. This imparts a greater Damon–Eshbach character to the spin wave, leading to an increased wavelength. Using the LLG-based model to fit the measured field-dependence of the spin-wave length (**Fig. 4c** and **Methods)**, we extract an anisotropy field of $B_K$ = 1 mT. The small discrepancies with the data can be attributed to the infinite-plane approximation used in the model.

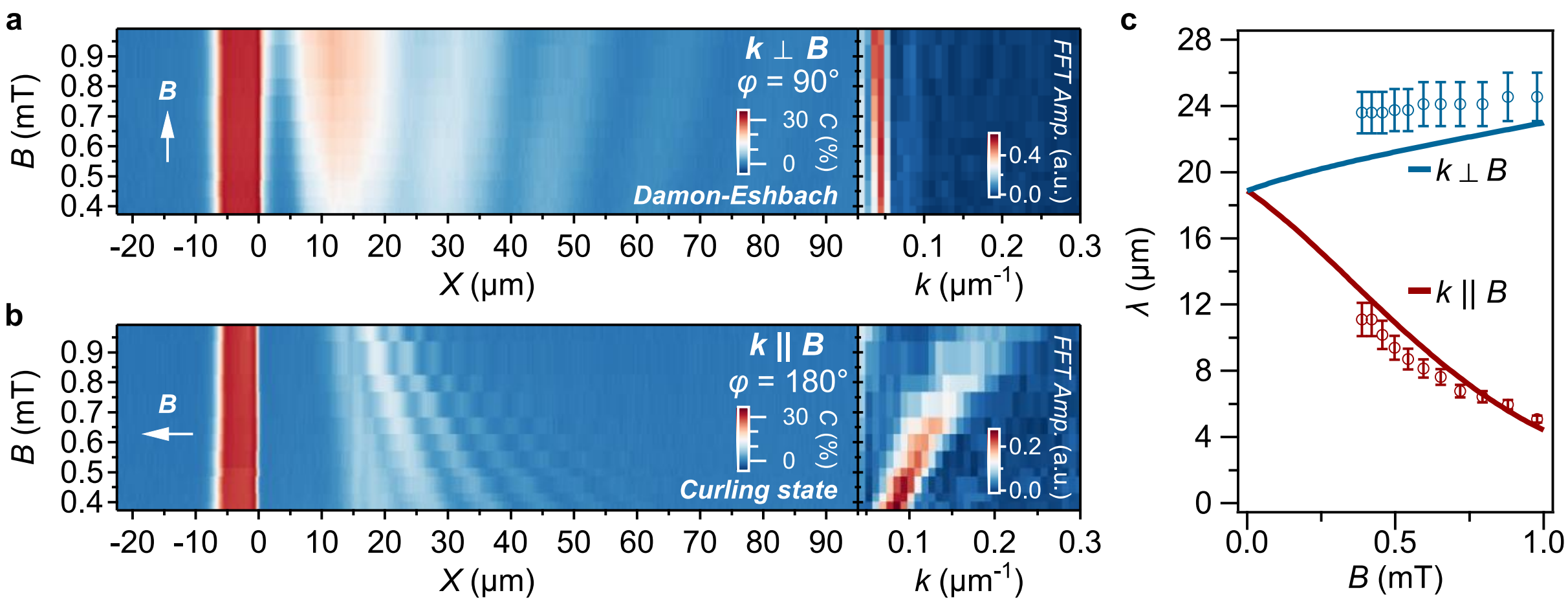

**Figure 4. Low-field control of the dispersion of the spin-wave length at the magnetic edge. a-b)** Spin-wave maps *vs.* the magnitude of an applied magnetic field oriented perpendicular (**a**) and parallel (**b**) to the spin-wave vector (left panels), with the corresponding wavenumbers extracted via fast Fourier transformation (right panels). The direction of the applied field is indicated by white arrows. **c**) Spin-wave length *vs.* field strength extracted from **a** and **b** together with a fit (solid lines) based on the Landau-Lifshitz equation. From the fit, we extract $B_K$ = 1.0(2) mT. All corresponding spatial spin-wave maps are shown in **Supplementary Section 2**.

## Spin-wave propagation in bistable, inhomogeneous and field-rotation-dependent spin-textures

The combination of uniaxial anisotropy, Zeeman interaction, and shape anisotropy induced by magnetic edges can lead to inhomogeneous spin textures such as curling magnetizations or the formation of Néel domain walls.[34,43,44] In particular, close to the film edge and for applied fields perpendicular to the edge, such textures are prone to arise because of the competition between demagnetizing field, anisotropy and Zeeman energies.[34] We investigate how this competition leads to bistable, field-rotation-dependent spin textures and how these textures can be used to realize different regimes of local spin-wave transport (**Fig. 5)**.

To do so, we measure (**Fig. 5.a**) an isofrequency spin-wave map for a field applied perpendicular to the microstrip ($\varphi = 0°$) after initializing the magnetization with a 20 mT field at either $\varphi = 45°$ (left panel) or at $\varphi = -45°$ (right panel), representing a counterclockwise and clockwise rotation respectively. In both cases, the spin wave is clearly visible beyond $X \approx 5$ μm, but with a remarkably different profile: For counterclockwise rotation, the ESR contrast varies smoothly, while in the clockwise case there is a sharp contrast peak. In both cases, the wavelength of the spin wave then changes gradually with further increasing distance to the edge. This dependence on the magnetization history is robust over several rotations and observed at different field strengths (**Supplementary Section 2**).

We attribute this history-dependent phenomenology to different inhomogeneous spin textures underlying the spin-wave transport, with the small in-plane anisotropy playing a crucial role in breaking the symmetry of the system. To characterize the spin textures, we measure the static, out-of-plane field component by spatially mapping the NV ESR spectrum as a function of $x$ (**Supplementary Section 3**) and extract the out-of-plane component of the magnetic field (**Fig. 5b**), observing fields up to 3 mT at the film edge. These ESR maps do not vary significantly over the $y$ direction, indicating a translationally invariant system along $y$.

The smoothly changing static-field profile observed for the counterclockwise rotation (left panel, **Fig. 5b**) can be well fitted by a curling magnetization profile,[45] yielding a curling length of 19(3) μm (see **Supplementary Section 1.5**). Such curling arises from the competition between the Zeeman energy, which favors aligning the spins to the applied field, and the demagnetizing energy cost of spins pointing perpendicularly to the film edge. The curling length depends on both the film thickness and the saturation magnetization (Supplementary Section 1.5), which can be employed as experimental parameters for tailoring the curling profile and, consequently, engineering spin-wave propagation. On the other hand, explaining the doubly-peaked static-field profile of the clockwise rotation **(Fig. 5c)** requires micromagnetic simulations. Using the in-plane anisotropy field extracted above, we reproduce both situations using Mumax3[46] (**Fig. 5c, Supplementary Section 1.6)**: For the counterclockwise rotation case (left panel **Fig. 5c**), we find a curling spin texture (see arrows), whereas the counter-clockwise case (right panel **Fig. 5c**) yields a Néel domain wall in addition to a curling spin texture (sharp switch between purple to green region). The out-of-plane magnetic field generated by these two spin textures closely agrees with our measurements (**Fig. 5c**). In our calculations, the position of the domain and the relative field intensity between the doubly-peaked static-field profile depend sensitively on the anisotropy constants and, to a lesser extent, on the saturation magnetization. Nonetheless, across all tested parameter sets, the formation of a domain wall remains a robust feature, as shown in the **Supplementary Section 1.6**.

Finally, we study the spin textures at the film edge for other in-plane field angles. When the field is along the microstrip (φ = 90° and φ = 270°), we measure a spatially constant ESR frequency (**Supplementary Section 3.1)**. For intermediate angles (as φ = 45° or φ = -45°) we always observe a curling magnetization and no domain wall, independently of the history of the magnetic field (**Supplementary Section 3.3)**. Overall, the magnetic anisotropy enables the controlled generation of different inhomogeneous spin textures by choosing the handedness of the in-plane field rotation.

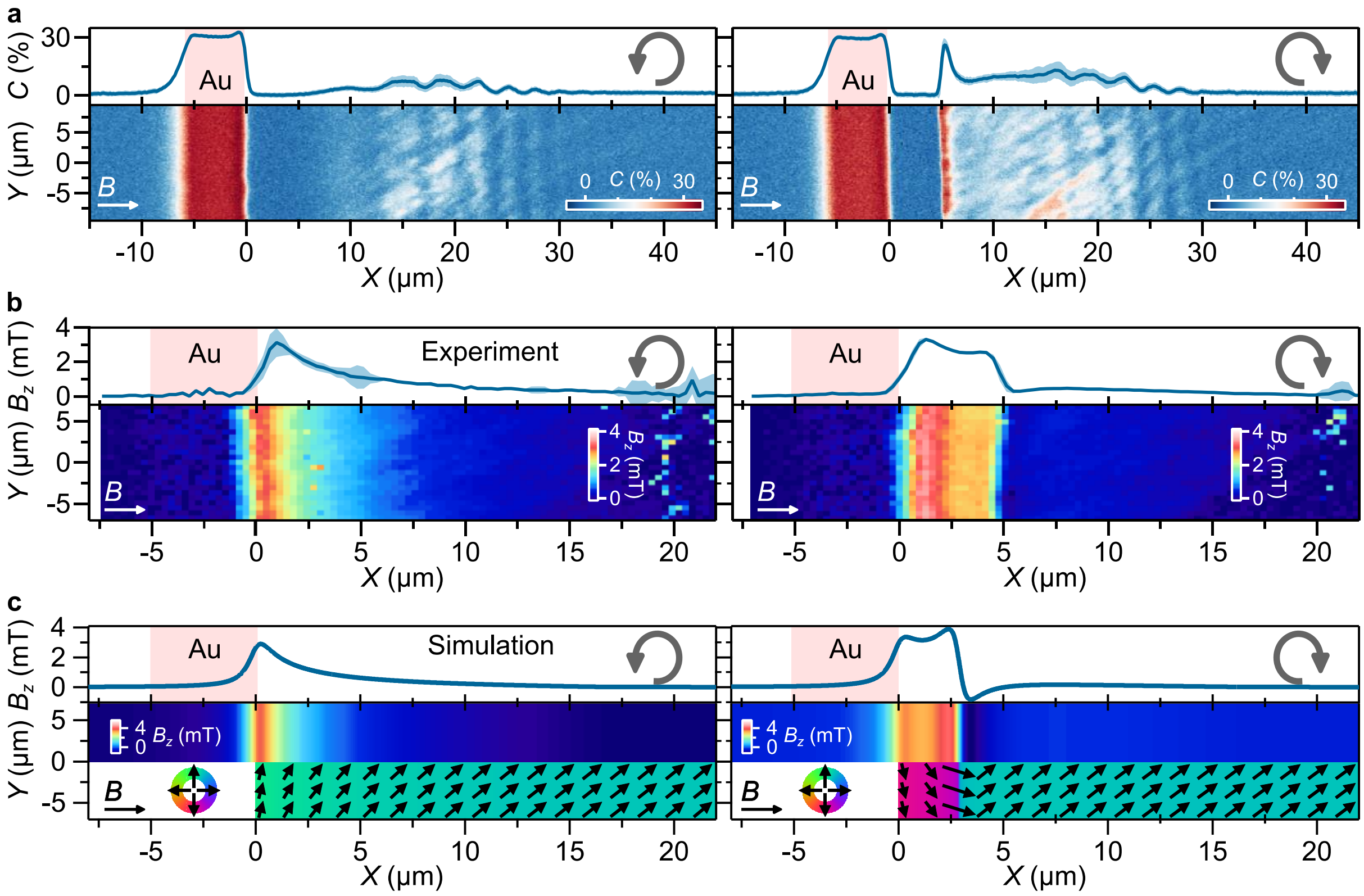

**Figure 5.- Spin-wave imaging in bistable, inhomogeneous spin textures created by magnetic-history control**. **a**) Spin-wave maps and their *Y*-averages at $B = 0.98$ mT and $f_{NV} = 2.87$ GHz with the bias field along *X* (white arrows). The left (right) panels are obtained after rotating the bias field counterclockwise (clockwise), as indicated by the grey arrows. **b**) Two-dimensional maps of $B_z$, showing translational invariance along *Y*, with their *Y*-averaged profile. The counterclockwise (clockwise) case shows a single (double) peak in $B_z$. **c**) Micromagnetic calculations of the spatial spin textures for counterclockwise (left panel) and clockwise (right panel) rotation of the applied field, showing curling spin textures and the deterministic, field-rotation controlled nucleation of a domain wall (bottom panel). The computed out-of-plane field generated by the spin textures is shown in the top panels. Red area in top panels indicate the location of the gold (Au) microstrip.

**Discussion**

The central concept of the spin-based spin-wave imaging approach we introduced here is the use of sensor spins that have an intrinsic anisotropy axis perpendicular to the magnetic film: this geometry enables decoupling to first order the magnetic-field control of the spin-wave spectrum, which is sensitive to in-plane fields, from that of the spin-resonance frequencies, which are sensitive to out-of-plane fields, to first order. This approach is general and should be applicable for quantum sensing of other systems with intrinsic anisotropy axes.

We used this approach for isofrequency imaging of field-controlled spin waves in a permalloy magnetic film, using both NV and $V_B$ centers to realize different detection frequencies. Other frequencies can be accessed by controlling the sensor spins using fields along the spin anisotropy axis, or by selecting other optically active spin-defects such as silicon vacancies in silicon carbide,[31] different spin defects in boron nitride (including those with in-plane anisotropy axes),[30,47] or molecular color centers.[48]

We used the ability to image both static and dynamic magnetic fields to non-reciprocally nucleate and image target spin textures and the resulting spin-wave transport at the edge of magnetic half-plane used as a device model. From a fundamental point of view, being able to image both spin waves and the underlying spin textures could reveal the interplay between coherent spin waves and other non-trivial spin textures, such as skyrmions or hopfions.[49,50] Regarding applications, we envision anisotropy-engineering as a key parameter towards the development of spin wave devices, motivating the development of selective film growth techniques for tailoring not only the strength but the angle of the magnetic anisotropy with respect the spin wave propagation and allowing the design and optimization of devices working at a single-frequency.[44]

## Methods

*$V_B$ centers in hBN thin-layer*

Previously we determined that the combination of boron-10 and nitrogen-15 isotopes are the best combination for quantum sensing.[21] Isotopically enriched $h^{10}B^{15}N$ crystal flakes were grown by the atmospheric pressure high temperature (APHT) method, previously described in detail.[28,51] Briefly, the process starts by mixing high purity 98% enriched boron-10, with nickel and chromium with mass ratios of 3.72:48.14:48.14, respectively. The mixture is then heated at 200°C/h under 97% enriched nitrogen-15 and hydrogen gas at pressures of 787 and 60 torr respectively to 1550°C, to produce a homogeneous molten solution. After 24 hours, the solution is slowly cooled at 1°C/h, to 1500°C, then at 50°C/h to 1350°C, and 100°C/h to room temperature. The $h^{10}B^{15}N$ solubility decreases as temperature is reduced, causing crystals to precipitate on the surface of the metal. We subsequently exfoliated the $h^{10}B^{15}N$ flakes from the metal with thermal release tape. The free-standing $h^{10}B^{15}N$ crystalline flakes are typically more than 20 $\mu$m thick. To prepare the boron vacancies ($V_B$), the isotope enriched $h^{10}B^{15}N$ flakes were neutron irradiated at the Ohio State University for a cumulative fluence of $2.9 \times 10^{16}$ neutrons/cm$^2$. We then mechanically exfoliated the hBN flakes from the neutron-irradiated bulk crystals and transferred them on top of the film by conventional two-dimensional transfer techniques. The hBN flake shown in the main text has a thickness of 55 nm, as determined by atomic force microscopy (**Supplementary Fig. 1**).

*NV centers in diamond*

Single out-of-plane (111) NV centers in a diamond incorporated into a tuning fork are obtained from QZABRE (QST-Q7-S-111-OOP and QST-Q5-S-111-OOP, for the extended and half-plane film, respectively). Measurements are taken at a sensor-sample separation of ~500 nm (Py), unless mentioned otherwise, yielding a ~1 micrometer spatial resolution while filtering out the stray fields of smaller spin waves potentially generated by spin-wave scattering.[52] This resolution is similar to the optical-diffraction-limited resolution of our VB-measurements. Measurements at different sensor-sample separations are shown in the **Supplementary Section 2.2.5**.

*Quantum sensing protocol*

We focus a 515 nm laser (Cobolt 06-MLD) with an optical power of 40 μW onto the single NV and 360 μW onto the hBN flake and record the resultant photoluminescence using an avalanche photodiode (Laser Components COUNT-250n). A microwave source (Rohde & Schwarz SGS100A) is connected to the microstrip to generate spin waves and drive the ESR. The magnetic imaging is performed using a bimodal scheme:[12] For every pixel, we record the photoluminescence with the microwave on and off (dwell time of 50 μs), from which we calculate the ESR contrast $C$ as discussed in the main text. In the case of the NV and hBN data shown in **Fig. 2d**, the spatial oscillations arise due to interference between the microwave magnetic stray fields generated by the spin waves and the microwave field generated by the microstrip.[12] To enhance the contrast of these oscillations, and thereby speed up the measurements, we apply an additional, auxiliary microwave field of the same frequency for the hBN data shown in **Fig. 2c**, **Fig. 2e**, **Fig. 2h** and **Supplementary Section 2.1.2** using an antenna loop placed at ~60 μm from the measurement area. Extracting the decay length from the presented imaging results is challenging because it requires precise modelling of the local microwave magnetic field amplitude, which is a result of interference between the spin-wave stray fields and any additional magnetic fields resonant with the NV center. In addition to the direct field generated by the current in the microstrip, additional NV-resonant fields (such as those generated by inductively excited eddy currents in the metallic film) could be present. Spatial scanning is performed by placing the magnetic film on XYZ-piezoelectric stages.

Further details on the magnetic field control and data analysis (spin-wave length determination from FFT transforms and curling analysis) are given in **Supplementary Section 1.3-1.5.**

*Device fabrication*

Permalloy (NiFe 81%19% from Kurt J. Lesker Company, EVMNIFEEXE-D, 99.95% purity for the extended device; NiFe 80%20% from Neyco, 99.95 % purity for the half-plane device) films with 50 nm thicknesses are evaporated on top of 300 nm-thick $SiO_2$/(100)-Si substrates using thermal evaporation inside a vacuum chamber (Angstrom Engineering Nexdep installed inside a glovebox with a base pressure of $7\cdot10^{-7}$ mbar at a rate of 0.5 Å/s) or electron beam evaporation. For the extended films, a 5 nm layer of $SiO_2$ is coated from a sputtering source using a 2" $SiO_2$ target (RF power: 90 W; deposition rate: 0.1 Å/s; base pressure: $7\cdot10^{-7}$ mbar; deposition pressure: 66.6 mbar.). The $SiO_2$ thickness was calibrated using ellipsometry.

For the spin-wave excitation in the extended films, we fabricated a 3-mm-long and 10-μm-wide microstrip (5 nm titanium / 80 nm gold) using electron beam lithography (Raith Pioneer Two) with a double layer resist (A6 495K / A3 950K). For the half-plane device, we fabricated a 1-mm-long and 5-μm-wide microstrip (5 nm titanium / 205 nm gold; A6 495K / A3 950K, Py: A4 495K / A3 950K) at 375 nm from the half-plane edge (**Supplementary Fig. 2**), using in the fabrication process an additional top layer of Elektra92, followed by an O2-plasma descum and subsequent thin film deposition using electron beam evaporation. The morphology is characterized by atomic force microscopy (**Supplementary Fig. 2**), with ~2.4 nm roughness (RMS, 11x18 μm2).

We characterized the extended Py films at room temperature using SQUID magnetometry (SQUID magnetometer MPMS-XL-7 from Quantum Design) under in-plane and out-of-plane fields (**Supplementary Fig. 3**), where the linear diamagnetic silicon magnetic response is subtracted for clarity. The observed small hysteresis is indicative of the presence of uniaxial magnetic anisotropy, as commonly present in permalloy films.[41,42]

*Micromagnetic simulations*

The magnetization of the film was simulated using the Mumax3 micromagnetics package.[46] Details are given in the **Supplementary Section 1.6.**

*First-order decoupling the sensor spin frequency from the spin-wave spectrum*

A central concept of our work is to decouple the magnetic-field-control of the sensor spin frequency from that of the spin-wave spectrum. Here we describe how this can be achieved by using a sensor spin with an anisotropy axis that is oriented perpendicular to the plane of the thin-film magnet. Although in this work we focused on a soft magnetic film with an out-of-plane hard axis determined by shape anisotropy, the sensing concept we introduce is general: by orienting the easy axis of the sensor spin along the hard axis of a magnetic system, we decouple control of the sensor spin frequency from that of the magnetic dynamics to first order. This approach enables isofrequency imaging of field-controlled spin waves as well as spin isowavelength imaging at different field-controlled sensor frequencies.

*Magnetic-field dependence of the sensor-spin ESR*

Color centers such as the NV center in diamond and the $V_B$ center in boron nitride have an effective spin of magnitude $S=1$.[1,21] As a result, the crystallographic symmetry of the defect endows the spin with an intrinsic anisotropy axis of which the orientation is defined by the crystallographic orientation of the defect. Here we show that this intrinsic anisotropy renders the electron spin resonance frequencies first-order (second-order) sensitive to magnetic fields that are oriented parallel (perpendicular) to the anisotropy axis.

The spin eigenfrequencies of the defect follow from diagonalizing the spin Hamiltonian

$$H = DS_z^2 + \gamma \mathbf{B} \cdot \mathbf{S} \quad (1)$$

where $\mathbf{S} = [S_x, S_y, S_z]$ are the Pauli spin-1 matrices, $\gamma$ is the electron gyromagnetic ratio, and we have neglected small terms such as interaction with strain. Diagonalizing this Hamiltonian leads to three eigenfrequencies $f_{\{1,2,3\}}$, from which we find the 'upper' and 'lower' ESR frequencies depicted in Fig.1, defined as $f_u = f_3 - f_1$, and $f_l = f_2 - f_1$. A measurement of $f_{\{l,u\}}$ then yields the magnitude $B$ of the magnetic field **B**:[2]

$$\gamma B = \frac{1}{\sqrt{3}}\sqrt{f_u^2 + f_l^2 - f_u f_l - D^2} \tag{2}$$

whereas the angle θ of **B** w.r.t the anisotropy axis follows from

$$\cos^2\theta = \frac{-(f_u+f_l)^3 + 3f_u^2 + 3f_l^2 + 2D^2}{27D\gamma^2B^2} + \frac{1}{3} \tag{3}$$

By writing $f_{u,l} = D + \delta_{u,l}$ and expanding the preceding two equations to 2nd order in $\delta_{u,l}$, we arrive at two equations from which we can solve for $\delta_{u,l}$ as a function of the applied field **B**:

$$D(\delta_u + \delta_l) - \delta_u^2 - \delta_l^2 + 4\delta_u\delta_l = 3\gamma^2B^2(1 - 3\cos^2\theta) \tag{4}$$

and

$$D(\delta_u + \delta_l) + \delta_u^2 + \delta_l^2 - \delta_u\delta_l = 3\gamma^2B^2 \tag{5}$$

We solve these equations for two cases: 1) If the field is oriented *along* the anisotropy axis (θ = 0), we find $\delta_u = \gamma B$ and $\delta_l = -\gamma B$. I.e., the ESR frequencies are *first-order sensitive* to B (see blue and red lines in **Fig. 1c**). 2) If, on the other hand, the field is *perpendicular* to the anisotropy axis (θ = π/2), we find

$$\delta_u = 2\frac{\gamma^2B^2}{D} \tag{6}$$

and

$$\delta_l = \frac{\gamma^2B^2}{D} \tag{7}$$

*I.e.*, the ESR frequencies are *decoupled* from the magnetic field B to first order (see blue and red lines in **Fig. 1d**), resembling a clock transition. This enables isofrequency imaging of field-controlled spin waves, as demonstrated in this work.

*Magnetic-field dependence of the spin-wave spectrum of a soft magnetic film*

Similarly, soft magnets such as permalloy films have an out-of-plane hard axis because of the demagnetizing energy cost of the out-of-plane magnetization configuration. Here we show that this intrinsic anisotropy renders the spin-wave spectrum first-order sensitive (insensitive) to in-plane (out-of-plane) magnetic fields, as illustrated in the color maps of **Fig. 1c,d**.

The free energy density $F$ of a homogeneously magnetized, soft magnetic film with saturation magnetization $M_s$ can be modeled by

$$\frac{F}{M_s} = \frac{\mu_0}{2} M_s \cos^2\theta_m - B\cos(\theta_m - \theta) \tag{8}$$

where $\theta_m$ and $\theta$ are the angle of the magnetization and of the applied field w.r.t the plane normal, respectively. For a field applied along the plane normal ($\theta$=0), minimizing this expression with respect to $\theta_m$ yields $\cos\theta_m = \frac{B}{\mu_0 M_s}$. I.e., the out-of-plane component of the magnetization increases linearly with $B$. In this work, as $\mu_0 M_s \approx 1$ T for permalloy and the applied fields are $B \approx 0.01$ T, the out-of-plane component remains very small, on the order of 1% .

To show how the dipolar spin-wave dispersion of the soft ferromagnetic film depends on the applied field angle, we express the free energy density as

$$\gamma\frac{F}{M_s} = -\omega_B b_i' m_i' - \frac{\omega_M}{2} m_i' \int \Gamma_{ij}'(\mathbf{r}-\mathbf{r}') m_j'(\mathbf{r}') d\mathbf{r}' \tag{9}$$

where we use Einstein summation over repeated indices, $\Gamma$ is the dipolar tensor, $\omega_B = \gamma B$, $\omega_M = \gamma\mu_0 M_s$, and $\mathbf{b}' = b_{x'}, b_{x'}, b_{z'}$ , and $\mathbf{m}' = m_{x'}, m_{x'}, m_{z'}$ are unit vectors specifying the direction of the applied magnetic field and the magnetization in the lab frame, respectively (note we use a prime (‘) to specify the lab frame).

To determine the spin-wave dispersion, we define a magnet frame (*x,y,z*), in which the equilibrium magnetization **m** points along *z*. The magnetizations in lab and magnet frames are related by $\mathbf{m}' = R\mathbf{m}$, where the rotation matrix *R* is

$$R(\theta_m, \phi_m) = \begin{bmatrix} \cos(\theta_m)\cos(\phi_m) & -\sin(\phi_m) & \sin(\theta_m)\cos(\phi_m) \\ \cos(\theta_m)\sin(\phi_m) & \cos(\phi_m) & \sin(\theta_m)\sin(\phi_m) \\ -\sin(\theta_m) & 0 & \cos(\theta_m) \end{bmatrix} \tag{10}$$

so that $\theta_m$ and $\phi_m$ are the zenith and azimuthal angles of the equilibrium magnetization in the lab frame, respectively. As we do not include any in-plane magnetic anisotropy, we can set both the azimuthal angle of the applied field and magnetization to $\phi = \phi_m = 0$. Furthermore, we assume $\omega_B \ll \omega_M$, which is the case in this work, as $\mu_0 M_s \approx 1$ T for permalloy and the applied fields are $B \approx 0.01$ T.

We find the spin-wave dispersion from the LLG equation, which is (without the damping term):

$$\dot{\mathbf{m}} = -\gamma \mathbf{m} \times \mathbf{B_{eff}} \tag{11}$$

The effective field $\mathbf{B_{eff}}$ should be calculated in the magnet frame. Its components are given by

$$\gamma B_n = -\frac{\gamma}{M_s}\frac{\partial F}{\partial m_n} = \omega_B b_i' R_{in} + \omega_M \int R_{in}\Gamma_{ij}'(\mathbf{r}-\mathbf{r}')R_{jl}m_l(\mathbf{r}')d\mathbf{r}' \tag{12}$$

Taking the Fourier transform over (*x,y*) and averaging the dipolar field over the film thickness *t* (to get the coupling to the lowest-order perpendicular spin-wave mode), we get

$$\gamma B_n = \omega_B b_i' R_{in} + \omega_M \bar{\Gamma}_{nl}(\mathbf{k}) m_l(\boldsymbol{k}) \tag{13}$$

where **k** is the in-plane spin-wave vector and $\bar{\Gamma}_{nl}(\mathbf{k}) = \mathrm{R_{in}}\bar{\Gamma}'_{\mathrm{ij}}(\mathbf{k})\mathrm{R_{jl}}$, with

$$\bar{\Gamma}'(\mathbf{k}) = -\begin{bmatrix} f_t \cos^2\phi_k & f_t \sin\phi_k \cos\phi_k & 0 \\ f_t \sin\phi_k \cos\phi_k & f_t \sin^2\phi_k & 0 \\ 0 & 0 & 1-f_t \end{bmatrix} \tag{14}$$

where $f_t \approx \frac{kt}{2}$ because $kt \ll 1$ for the dipolar spin waves considered in this work, and $\phi_k$ is the in-plane angle of the spin-wave vector.

When the field is oriented out of plane ($\mathbf{b}' = (0,0,1)$ and $\theta = 0$), the rotation matrix becomes

$$R = \begin{bmatrix} \cos(\theta_m) & 0 & \sin(\theta_m) \\ 0 & 1 & 0 \\ -\sin(\theta_m) & 0 & \cos(\theta_m) \end{bmatrix} \tag{15}$$

where $\cos(\theta_m) = \frac{\omega_B}{\omega_M}$ is the equilibrium orientation of the magnetization as derived above. We find

$$\gamma B_z = \omega_B \cos(\theta_m) + \omega_M \bar{\Gamma}_{zz}(\mathbf{k} = \mathbf{0}) \tag{16}$$

$$\gamma B_x = \omega_M \left( \bar{\Gamma}_{xx} m_x + c \bar{\Gamma}_{xy} m_y \right) \tag{17}$$

$$\gamma B_y = \omega_M \left( \bar{\Gamma}_{yx} m_x + \bar{\Gamma}_{yy} m_y \right) \tag{18}$$

where $B_z$ is the *static* part of the field along $z$ and $B_{x,y}$ are the *dynamic* parts of the transverse fields, needed for formulating the linearized LLG equation. Using $\bar{\Gamma}_{nl}(\mathbf{k}) = \mathrm{R_{in}} \bar{\Gamma}'_{\mathrm{ij}}(\mathbf{k}) \mathrm{R_{jl}}$, we have:

$$\bar{\Gamma}_{xx} = -\left( \frac{\omega_B^2}{\omega_M^2} \left( f_t (1 + \cos^2 \phi_k) - 1 \right) + 1 - f_t \right) \tag{19}$$

$$\bar{\Gamma}_{xy} = \bar{\Gamma}_{yx} = -\frac{\omega_B}{\omega_M} f_t \sin \phi_k \cos \phi_k \tag{20}$$

$$\bar{\Gamma}_{yy} = -f_t \sin^2 \phi_k \tag{21}$$

$$\bar{\Gamma}_{zz}(\mathbf{k} = 0) = -\frac{\omega_B^2}{\omega_M^2} \tag{22}$$

The LLG equation becomes:

$$\omega \begin{bmatrix} m_x \\ m_y \end{bmatrix} = i \begin{bmatrix} -\omega_1 & \omega_3 \\ -\omega_2 & \omega_1 \end{bmatrix} \begin{bmatrix} m_x \\ m_y \end{bmatrix} \tag{23}$$

where we defined

$$\omega_0 = \gamma B_z = \omega_B \cos(\theta_m) + \omega_M \bar{\Gamma}_{zz}(\mathbf{k} = \mathbf{0}) = 0 \tag{24}$$

$$\omega_1 = \omega_M \bar{\Gamma}_{xy} = -\omega_B f_t \sin \phi_k \cos \phi_k \tag{25}$$

$$\omega_2 = \omega_0 - \omega_M \bar{\Gamma}_{xx} = \omega_M \left( \frac{\omega_B^2}{\omega_M^2} (f_t + f_t \cos^2 \phi_k - 1) + 1 - f_t \right) \quad (26)$$

$$\omega_3 = \omega_0 - \omega_M \bar{\Gamma}_{yy} = \omega_M f_t \sin^2 \phi_k \quad (27)$$

The dispersion is given by the eigenvalues of the LLG equation:

$$\omega_{sw}^2 = \omega_2 \omega_3 - \omega_1^2 \quad (28)$$

$$\approx (\omega_M^2 - \omega_B^2) \sin^2 \phi_k \, f_t \quad (29)$$

where the approximation holds when $kt \ll 1$ such that we can neglect terms scaling with $f_t^2$. From this last expression, we see that for $\phi_k = \frac{\pi}{2}$ (Damon Eshbach waves), we get

$$f_t \approx \frac{\omega_{sw}^2}{\omega_M^2} \left( 1 + \frac{\omega_B^2}{\omega_M^2} \right) \quad (30)$$

which shows that the wavenumber of a spin wave with frequency $\omega_{sw}$ is *first-order insensitive* to the magnetic field $\omega_B$. This insensitivity is illustrated in **Fig. 1c**.

If, on the other hand, the magnetic field is applied *in-plane*, such that $\theta = \theta_m = \pi/2$, we have

$$\bar{\Gamma}_{xx} = -(1 - f_t) \quad (31)$$

$$\bar{\Gamma}_{xy} = \bar{\Gamma}_{yx} = 0 \quad (32)$$

$$\bar{\Gamma}_{yy} = -f_t \sin^2 \phi_k \quad (33)$$

$$\bar{\Gamma}_{zz}(\mathbf{k} = 0) = 0 \quad (34)$$

so that the spin-wave dispersion is

$$\omega_{sw}^2 = \left( \omega_B + \omega_M (1 - f_t) \right) (\omega_B + \omega_M f_t \sin^2 \phi_k) \quad (35)$$

We conclude that, for an in-plane applied magnetic field, the spin-wave number of the Damon Eshbach spin waves is *first-order sensitive* to the applied field:

$$f_t \approx \frac{\omega_{sw}^2}{\omega_M^2} - \frac{\omega_B}{\omega_M} \quad (36)$$

*Low-field spin-wave dispersion of a ferromagnetic film with a small in-plane anisotropy*

In **Fig. 3** (**Fig. 4**), we demonstrated isofrequency imaging of spin waves in the permalloy film as a function of the in-plane angle (magnitude) of an applied magnetic field. Here we describe the model used to analyze the observed field dependence of the spin-wave length in **Fig. 3** and **Fig. 4**.

Including an in-plane anisotropy of magnitude $\gamma B_K = \omega_K$ oriented along *y*, the free energy density is

$$\gamma \frac{F}{M_s} = -\frac{\omega_K^2}{2} m_{y'} - \omega_B b_i' m_i' - \frac{\omega_M}{2} m_i' \int \Gamma_{ij}'(\mathbf{r}-\mathbf{r}') m_j'(\mathbf{r}') d\mathbf{r}' \tag{37}$$

When the applied magnetic field is in-plane as defined by $\mathbf{b}' = (\cos\phi, \sin\phi, 0)$, the equilibrium magnetization lies in-plane, with an in-plane angle that follows from minimizing *F* with respect to $\phi_m$, yielding

$$\frac{\omega_K}{2} \sin 2\phi_m = -\omega_B \sin(\phi - \phi_m) \tag{38}$$

This equation needs to be solved numerically except when $\phi = \frac{\pi}{2}$ or $\phi = 0$, in which cases we find $\phi_m = \frac{\pi}{2}$ and $\cos\phi = \min\left(\frac{\omega_B}{\omega_K}, 1\right)$ respectively. Importantly, the magnetization angle makes a jump when the field crosses the hard axis, which manifests itself in the jump observed in **Fig. 3e-f**.

Following the same procedure as in the previous section, we find the spin-wave dispersion from the LLG equation:

$$\omega_{sw}^2 = [\omega_K \sin^2\phi + \omega_B \cos(\phi - \phi_m) + \omega_M(1 - f_t)] \cdot [-\omega_K \cos 2\phi + \omega_B \cos(\phi - \phi_m) + \omega_M f_t \sin^2(\phi_k - \phi)] \tag{39}$$

When $\omega_K, \omega_B \ll \omega_M$, solving for the spin-wave number $f_t \approx \frac{kt}{2}$ yields

$$f_t \approx \frac{\omega_{sw}^2}{\omega_M^2} - \frac{\omega_K}{\omega_M} - \frac{\omega_B}{\omega_M} \quad \text{for} \quad \phi_B = \phi = \frac{\pi}{2} \tag{40}$$

$$f_t \approx \frac{\omega_{sw}^2 - \omega_B^2}{\omega_M^2\left(1 - \frac{\omega_B^2}{\omega_K^2}\right)} \quad \text{for} \quad \phi_B = 0, \cos\phi = \frac{\omega_B}{\omega_K} \tag{41}$$

Similarly to before, the spin wavenumber scales linearly with field if the field is oriented along the easy axis ($\phi_B = \frac{\pi}{2}$), whereas if a small field ($\omega_B \ll \omega_K$) is applied perpendicular to the easy axis ($\phi_B = 0$), the spin-wave number is insensitive to the applied field and scales quadratically. If on the other hand the applied field becomes *comparable* to the anisotropy field ($\omega_B \approx \omega_K$), the spin-wave number changes rapidly with field, as can be seen from the last equation. This is demonstrated in **Fig. 4**. This shows how small intrinsic anisotropies enable tuning the sensitivity of the spin-wave lengths to a magnetic control field.

**Data Availability**

Data is publicly available at Zenodo (https://doi.org/10.5281/zenodo.17612464).

**Acknowledgements**

This research is supported by The Kavli Foundation and the Dutch Research Council (NWO) under awards VI.Vidi.193.077, NGF.1582.22.018. S.M.-V. acknowledges the support from the European Commission for a Marie Sklodowska–Curie individual fellowship No. 101103355 - SPIN-2D-LIGHT and the Synergy postdoctoral grant by the Kavli Institute of Nanoscience Delft thanks to The Kavli Foundation. S.K. acknowledges support from NWO via VI.Veni.222.296. Support for the $h^{10}B^{15}$ crystal growth was provided by the United States National Science Foundation and the Agence Nationale de la Recherche, award number 2413808 (ANR/NSF/QISE). Neutron irradiation of the $h^{10}B^{15}N$ was supported by the U.S. Department of Energy, Office of Nuclear Energy under DOE Idaho Operations Office Contract DE-AC07-05ID14517 as part of Nuclear Science User Facilities award # 24-4911.

**Author Contributions Statement**

Conceptualization: S.M.-V. and T.S.; Sample design and fabrication: S.K. and S.M.-V.; hBN crystal growth and characterization of $V_B$ centers: T.P., J.H.E. and V.J.; hBN exfoliation and flake placement: S.M.-V.; Constructing experimental setup: S.M.-V. and T.S.; Measurements and software: S.M.-V. and Y.D.; Magnetic field calibration: Y.D., M.B. and S.M.-V.; Data analysis: S.M.-V., Y.D. and T.S.; Theory: T.S. and Y.M.B.; Micromagnetic simulations: A.B. and Y.M.B.; Figures: S.M.-V. and Y.D.; Funding acquisition: S.M.-V. and T.S.; Project administration: S.M.-V. and T.S.; Supervision: S.M.-V. and T.S.; Writing – original draft: S.M.-V., Y.D. and T.S.; Writing – review and editing: S.M.-V., Y.D., T.S., V.J. and Y.M.B. The manuscript was written through contributions of all authors. All authors have given approval to the final version of the manuscript.

**Competing Interests Statement**

The authors declare no competing interests.

## Figure Captions

**Figure 1.- Spin wave imaging using solid-state spin sensors that are first-order decoupled from the spin-wave spectrum. a, b)** Atomic structure of the nitrogen-vacancy (NV) center in the diamond carbon (C) lattice (a) and of the boron vacancy ($V_B$) in a hexagonal boron nitride layer (b). **c, d)** Dispersion of the spin sensors and the spin waves. Blue and red lines: Magnetic field dependence of the electron spin resonance (ESR) frequencies of the NV (zero-field splitting $D$ = 2.87 GHz) and $V_B$ (D = 3.5 GHz) center, respectively. Color map: spin-wave length λ. The red, blue and green arrows above the graphs indicate the spin anisotropy axis, externally applied magnetic field, and easy-plane magnetization respectively. Grey dashed line: the ferromagnetic resonance (FMR) according to Kittel's law. In (c), the magnetic field $B$ is applied *along* the spin anisotropy axis, linearly splitting the ESR frequencies $f_-$ and $f_+$ while leaving the spin-wave length λ unaffected to first order. In (d), $B$ is applied *perpendicular* to the spin anisotropy axis, rendering the ESR frequencies decoupled from $B$ to first order while strongly coupling to the spin-wave length. See **Methods** for the mathematical derivation. Inset in (d): sketch illustrating the spin-wave length λ. **e-f)** Optically detected ESR spectrum near zero applied magnetic field for the NV (e) and $V_B$ (f) spins. Monitoring the ESR contrast, defined as $C = 1 - PL_{MWon}/PL_{MWoff}$, where $PL_{MWon}$ ($PL_{MWoff}$) is the photoluminescence (PL) under (in the absence of) microwave driving, enables spin wave detection. The four-peak structure of the $V_B$ spectrum is associated to hyperfine coupling to nearby $^{15}N$ spins.[21]

**Figure 2.- Isofrequency imaging of field-controlled spin waves using color centers. a)** Sketch of the experiment: a microwave (MW) current in a 80-nm-thick gold (Au) microstrip excites spin waves (green arrows) in a 55-nm-thick permalloy (Py) film (**Methods**). We detect the waves using both an ensemble of $V_B$ sensor spins in a hexagonal boron nitride flake placed on top of the sample and a single NV sensor spin in a diamond tip scanned across the sample. **b**) Optical image showing the $V_B$ -containing boron-nitride flake on top of the Py sample and Au microstrip. Scale bar: 10 µm. **c**) ESR contrast map showing spin waves travelling perpendicular (Damon-Eshbach spin waves) to the microstrip with wavevector **k**. A wavelength of 7.9(8) µm is extracted by fast Fourier transformation (FFT, **Supplementary Section 1.4**). An in-plane field $B$ = 1.6 mT along the microstrip and a $V_B$ -resonant drive frequency of 3.52 GHz yields Damon-Eshbach spin waves. Scale bar: 10 µm. **d-e**) Spin-wave maps (bottom panel) and their average along Y (top panel) for $B$ = 1.6 mT applied in-plane along the microstrip (white arrows). The drive frequency is 3.55 GHz for the $V_B$ measurement in (d) and 2.87 GHz for the NV measurement in (e). The FFT-extracted wavelengths are 8.2(8) and 10.6(9) µm, respectively. The gold microstrip is located between X ∈ [-5, 5] µm. Shaded area in top panels: ±1 standard deviation. **f-g**) Isofrequency imaging of field-controlled spin waves. Measured (left panels) and calculated (right panels) 1D spin-wave maps vs. $B$. The field $B$ is applied in-plane along the microstrip. Drive frequency: 3.44 GHz for the $V_B$ measurement in (f) and 2.87 GHz for the NV measurement in (g). An external MW reference field is added in f (see **Methods**). **h)** Spin-wave length vs. field strength $B$. The 1D profiles shown in panels (f) and (g) are obtained from averaging the 2D spatial contrast maps presented in **Supplementary Section 2.1**. The spin-wave length shown in (h) is extracted using a fast Fourier transform (FFT) applied to these 2D maps, as described in **Supplementary Section 1.4**.

**Figure 3. Isofrequency imaging of anisotropy-governed spin waves under angular field control. a)** Sketch of the experiment. A microwave (MW) current in a gold (Au, yellow) microstrip excites spin waves (green arrows) with wavevector k (white arrow) in a 50-nm-thick permalloy (Py) half-plane. A single NV spin (lattice structure shown in inset) in a scanning diamond tip images spin waves resonant with the NV ESR frequency by detecting their microwave magnetic stray fields. **b)** Optical image of the magnetic half-plane. Scale bar: 2 mm. **c)** Spatial spin-wave maps (bottom panels) and their average along the vertical direction (top panels) for eight different angles φ (in-plane angle from *x* towards *y* as defined in a) of the applied magnetic field ($B$ = 0.98 mT). Scale bar: 10 µm. **d)** Spin-wave maps *vs.* field angle φ (left panel) and the corresponding spin-wave number *k* extracted via fast Fourier transformation (FFT, right panel) at $B$ = 0.98 mT. **e-f)** Spin wave length *vs.* in-plane field angle φ at $B$ = 0.98 mT (**e**) and $B$ = 0.39 mT (**f**). The sudden change in (f) signifies the angle at which the applied field is perpendicular to the anisotropy axis, enabling us to extract an anisotropy angle of $\varphi_K$ = 73(2)° (highlighted with pink dashed line). Red lines: calculation based on the Landau-Lifshitz equation for an anisotropy field of $B_K$ =1 mT that is oriented at $\varphi_K$ = 73°. The corresponding spatial spin-wave maps are shown in **Supplementary Section 2.2**.

**Figure 4. Low-field control of the dispersion of the spin-wave length at the magnetic edge. a-b)** Spin-wave maps *vs.* the magnitude of an applied magnetic field oriented perpendicular (**a**) and parallel (**b**) to the spin-wave vector (left panels), with the corresponding wavenumbers extracted via fast Fourier transformation (right panels). The direction of the applied field is indicated by white arrows. **c**) Spin-wave length *vs.* field strength extracted from **a** and **b** together with a fit (solid lines) based on the Landau-Lifshitz equation. From the fit, we extract $B_K$ = 1.0(2) mT. All corresponding spatial spin-wave maps are shown in **Supplementary Section 2**.

**Figure 5.- Spin-wave imaging in bistable, inhomogeneous spin textures created by magnetic-history control**. **a**) Spin-wave maps and their *Y*-averages at $B$ = 0.98 mT and $f_{NV}$ = 2.87 GHz with the bias field along *X* (white arrows). The left (right) panels are obtained after rotating the bias field counterclockwise (clockwise), as indicated by the grey arrows. **b**) Two-dimensional maps of $B_z$, showing translational invariance along *Y,* with their *Y*-averaged profile. The counterclockwise (clockwise) case shows a single (double) peak in $B_z$. **c)** Micromagnetic calculations of the spatial spin textures for counterclockwise (left panel) and clockwise (right panel) rotation of the applied field, showing curling spin textures and the deterministic, field-rotation controlled nucleation of a domain wall (bottom panel). The computed out-of-plane field generated by the spin textures is shown in the top panels. Red area in top panels indicate the location of the gold (Au) microstrip.